\documentclass[preprint2]{aastex63}
\hypersetup{linkcolor=blue,citecolor=blue,filecolor=cyan,urlcolor=magenta}
\usepackage{graphicx}
\usepackage{float}
\usepackage{latexsym,epsfig,amssymb,amsmath}
\usepackage[section]{placeins}
\usepackage{color}

\received{\today}
\revised{\today}
\accepted{TBD}
\submitjournal{AJ}

\graphicspath{{./}{figures/}}
\begin{document}

\title{Identifying acoustic wave sources on the Sun\\
I. Two-dimensional waves in a simulated photosphere}

\correspondingauthor{Shah Mohammad Bahauddin}
\email{shahmohammad.bahauddin@colorado.edu}

\author{Shah Mohammad Bahauddin}
\affiliation{Laboratory for Atmospheric and Space Physics, University of Colorado, Boulder, CO 80303, USA}
\affiliation{DKIST Ambassador, National Solar Observatory, Boulder, CO 80303, USA}

\author{Mark Peter Rast}
\affiliation{Department of Astrophysical and Planetary Sciences, University of Colorado, Boulder, CO 80309, USA}
\affiliation{Laboratory for Atmospheric and Space Physics, University of Colorado, Boulder, CO 80303, USA}



\begin{abstract}

The solar acoustic oscillations are likely stochastically excited by convective dynamics in the solar photosphere, though few direct observations of individual source events have been made and their detailed characteristics are still unknown.  Wave source identification requires measurements that can reliably discriminate the local wave signal from the background convective motions and resonant modal power.  This is quite challenging as these 'noise' contributions have amplitudes several orders of magnitude greater than the sources and the propagating wave fields they induce. In this paper, we employ a high-temporal-frequency filter to identify sites of acoustic emission in a radiative magnetohydrodynamic simulation. The properties of the filter were determined from a convolutional neural network trained to identify the two-dimensional acoustic Green's function response of the atmosphere, but once defined, it can be directly applied to an image time series to extract the signal of local wave excitation, bypassing the need for the original neural network.  Using the filter developed, we have uncovered previously unknown properties of the acoustic emission process. In the simulation, acoustic events are found to be clustered at mesogranular scales, with peak emission quite deep, about 500 km below the photosphere, and sites of very strong emission can result from the interaction of  two supersonic downflows that merge at that depth.  We suggest that the method developed, when applied to high-resolution high-cadence observations, such as those forthcoming with Daniel K. Inouye Solar Telescope (DKIST), will have important applications in chromospheric wave-studies and may lead to new investigations in high-resolution local-helioseismology.

\end{abstract}



\section{Introduction} \label{sec:intro}

The Sun and many stars are pulsationally stable but display acoustic oscillations none-the-less.  These stars are likely stochastically excited by small-scale convective dynamics, but the detailed properties of the acoustic sources are unknown.  Theoretical models differ, and observations are yet unable to differentiate between them.  

Understanding the sources of the solar acoustic oscillations is important in assessing their contributions to observed oscillation spectra and consequently in using those spectra to determine stellar properties.  Global $p$-mode line shapes, and thus accurate frequency determinations~\citep[e.g.,][and references therein]{1993ApJ...410..829D, 1998ApJ...506L.147T, 2018ApJ...857..119B}, depend critically on the depth and properties of the wave sources~\citep{1992A&A...265..771G, 1993A&A...274..935G, 1995MNRAS.272..850R, 1996ApJ...472..882A, 1998ApJ...496..527R, 2020A&A...635A..81P}.  Moreover, direct contributions of the excitation events to the observations introduces a correlated noise component to the $p$-mode spectra~\citep{1997MNRAS.292L..33R, 1998ApJ...495L.115N}, which can reverse the line asymmetries~\citep{1993ApJ...410..829D} and be  used to determine the phase relationship between intensity and velocity fluctuations during excitation events~\citep[][though cf.,~\citealp{2020arXiv201102439P}]{2000ApJ...535..464S, 2001ApJ...561..444S, 2003ApJ...596L.117J}.  Local helioseismological deductions are similarly sensitive to the phase relationship between the waves and their source.   For example, the travel-time kernels used in time-distance helioseismology depend on the assumptions about the source characteristics~\citep{2002ApJ...571..966G, 2004ApJ...608..580B}, and source properties may be particularly critical in the interpretation of multi-height local helioseismological measurements if the source is spatially and temporally extended, as it is likely to be.

Stochastic excitation by turbulent convection can result from several processes.  Approximately monopolar, dipolar, and quadrupolar emission results from fluid compression (volumetric changes), buoyant acceleration in a stratified medium (external stresses), and the divergence of the fluctuating Reynolds stresses (internal stresses) respectively~\citep[e.g.,][]{1990ApJ...363..694G,1999ApJ...524..462R}.  Early studies focused on quadrupolar excitation by
turbulent motions, the Lighthill mechanism~\citep{1952RSPSA.211..564L, 1954RSPSA.222....1L, 1967SoPh....2..385S, 1977ApJ...212..243G, 1990ApJ...363..694G,1992MNRAS.255..639B}, which scales as a high power of the turbulent flow Mach number.  This mechanism may be most readily observe on the Sun within intergranular lanes in the deep photosphere, as it is there that the flow is most turbulent, with the granular flow otherwise highly laminarized by the steep photospheric stratification~\citep[e.g.][]{1997A&A...328..229N}. There is some modeling~\citep[e.g.][]{2000ApJ...541..468S} and some observational~\citep[via acoustic flux measurements,][]{1995ApJ...444L.119R,1998ApJ...495L..27G} evidence that solar acoustic excitation preferentially occurs in granular downflow lanes~\citep{1995ApJ...444L.119R,1998ApJ...495L..27G, 2000ApJ...541..468S, 2000ApJ...535.1000S}.  

The importance of monopolar and dipolar emission due to rapid local cooling (radiatively induced entropy fluctuations) and consequent buoyant acceleration of the fluid in the solar photosphere is also recognized~\citep{1991LNP...388..195S,1994ApJ...424..466G, 1997ASSL..225..135R, 1998IAUS..185..199N, 1999ApJ...524..462R, 2001A&A...370..136S, 2001A&A...370..147S}, and the particular importance of granular fragmentation and the formation of new convective downdrafts in the solar photosphere has been emphasized~\citep{1993ApJ...408L..53R, 1995ApJ...443..863R}.  Direct observation of wave emission during granule fragmentation has been reported~\citep{2010ApJ...723L.175R, 2010ApJ...723L.134B}, and helioseismic phase difference spectra show a velocity/intensity phase relation consistent with downflow plume formation~\citep{1999ApJ...516..939S, 2000ApJ...535..464S, 2001ApJ...561..444S}.  Finally, solar flares have been implicated as strong acoustic sources~\citep[e.g.,][] {1998Natur.393..317K, 2003SoPh..218..151A, 2005ApJ...630.1168D}, though their coupling and energetic importance to solar $p$-modes is only partially understood~\citep[][]{2014SoPh..289.1457L}.

It is likely that acoustic sources on the Sun leverage both turbulent pressure and entropy fluctuations, but the precise nature of the excitation events, their phasing and efficiency in coupling to the global modes, and thus their relative importance to excitation, has not yet been quantitatively determined.  
Regularly identification of individual acoustic sources that link the observed local wave field directly to a specific source site 
would advance this cause.  Additionally, detailed characterization of resolved sources could provide a basis for wave mode conversion studies and high-resolution local helioseismology employing the local wave field generated.  

The difficulties faced in resolving solar acoustic sources stem from the inherent challenges in separating the faint (three or more orders of magnitudes weaker than the background) local wave field induced by the acoustic events from the background superposition of granular motion and global resonant p-modes.  Simulations suffer a similar difficulty: the unambiguous separation of compressible convective motions from the contributions of individual wave sources to the total flow remains problematic.
While projection of a simulation solution onto resonant oscillation modes is readily achieved~\citep[e.g.,][]{1993ApJ...407..316B}, 
identification of the local wave response is difficult; while one can formally define the local wave field as distinct from 
the compressible convection, and thereby identify possible source mechanisms and how they depend on the non-wave flow~\citep[e.g.,][]{1999ApJ...524..462R}, unambiguous separation of these wave and non-wave components
is in general not possible because the radiated wave-field is not
viewed in a turbulence and source free region~\citep[see, for example, Lighthill's discussion of sound vs. pseudo-sound,][]{1962RSPSA.267..147L}.  Moreover, the very qualities that make the Sun an excellent resonant cavity also make it difficult to distinguish any individual local source.
Inevitably an episode of wave emission from a local source is a small component of the signal compared with the resonant accumulation of acoustic waves, and acoustic wave filters~\citep[e.g.,][]{1989ApJ...336..475T} act to also reduce the local source signature.  
It has thus proven extremely difficult, both observationally and theoretically, to disentangle source signatures from the background modal and convective motions by means of standard signal processing techniques, such as Fourier analysis. Since the spectral content of the acoustic sources overlaps that of the acoustic modes, and in part the granular motion, and since the amplitude of the signal is extremely weak, linear filtering and frequency domain noise reduction techniques most often fail in source detection.

In this paper, we report on a new robust method for the unambiguous identification of acoustic source sites in the photosphere of a MPS/University of Chicago Radiative MHD \citep[MURaM,][]{2005A&A...429..335V, 2009ApJ...691..640R, 2014ApJ...789..132R} magnetohydrodynamic simulation of the upper solar convection zone.  
The method was developed by first utilizing a deep learning algorithm to reliably identify the weak residual high-frequency signature of local acoustic sources in Doppler velocity maps and then deciphering what underlies its success.  We have diagnosed what the learning algorithm is detecting, mimicked the filter it is applying, and applied the filter directly to the simulated photospheric time series, bypassing the dependence on deep-learning and allowing direct visualization of the local wave pulses that propagate outward from the acoustic source sites. To be effective, the acoustic-source filter thus derived requires high cadence ($< 3$ seconds) and high spatial resolution ($< 50$ km) timeseries. Fortuitously, the observational capabilities required to apply the filter to real solar data are just now becoming available with the commissioning of the National Science Foundation's Daniel K. Inouye Solar Telescope (DKIST).    

\section{Building the acoustic-source filter} \label{sec:filter1}

\subsection{Convolutional neural network} \label{sec:cnn}
Neural networks are a class of algorithms that perform inference without using explicit instructions, relying on patterns and examples instead. They utilize computational statistics, in which algorithms build models based on nonlinear and nonparametric regression over sample data, known as ``training data,'' in order to make forecasts or decisions.  If the training data sufficiently broad to capture the relevant correlations, the network can then be used to make inferences within a domain of interest.  For our problem, we considered a network architecture inspired by the organization of the visual cortex, known as a convolutional neural network \citep[e.g.,][and references therein]{2015Natur.521..436L}. In comparison to a fully-connected neural network, this architecture displays superior performance in fitting and classifying datasets of image time-series.  Successive convolution allows the network to more reliably construct each layer of representation while utilizing a 
smaller number of parameters. This advantage is fully realized when dedicated graphical processing units (GPU) are employed because, while they are more limited in local memory, these multi-stream processors allow for fast parallel processing reducing the time required to train networks, allowing training over extremely large data sets.

We have constructed a convolutional neural network tailored to the identification of local sources of acoustic waves in the photosphere of a MURaM simulation.  We have been successful in identifying sources in time-series of the evolving Doppler velocity on the two-dimensional photospheric ($\tau =1$) plane, pressure perturbations on the same surface, and the evolving continuum intensity.   The neural network in all cases is able to capture the spatial and temporal dependencies in the image sequences that define an acoustic source event through the application of multiple convolutional filters.  In this paper we focus, for simplicity, on source identification using the photospheric Doppler velocity time-series, though the steps taken and conclusions drawn are common to all variables.  Details of the network architecture and the training parameters are discussed in Appendix~\ref{sec:appendixA}.  Here we summarize the simulation and training scheme employed.

The physical dimensions of the MURaM simulation employed are  $L_x \times L_y \times L_z = 6.144 \times 6.144 \times 4$  Mm$^3$, where $L_z$ is the vertical dimension, with gridding for uniform $16$ km resolution in all directions. The simulation extends for 1 hour of solar time with a time step of $2.0625$ s (1800 frames in total). The data cube thus has the native shape $1800 \times 384 \times 384 \times 256$. The top boundary of the simulation is located 1.7 Mm above the mean $\tau = 1$ level; the depth of the convecting portion of the layer is 2.3 Mm.  Horizontally periodic boundary conditions were  employed during the simulation, along with a semi-transparent upper boundary (closed for downflows and opened for upflows) and an open lower boundary (mass preserving). For reference, the simulation is a re-run of the setup O16b from \cite{2014ApJ...789..132R} with non-grey radiative transfer and a domain extended an additional 1.024 Mm upwards into the chromosphere.  From it we extracted the Doppler velocity at an optical depth of $\tau = 1$, yielding a reduced  $1800 \times 384 \times 384$ data-cube for our primary analysis, though additional heights in the domain were examined as well, as discussed in $\S$\ref{sec:result}.    

The MURaM photospheric time-series can be thought of as being composed of three intrinsic components:  convective motions, modal oscillations and the wave field produced by local sources. In order to train the neural network to identify local wave sources two things are needed:  a template of the expected source signature and a source free time series of the granulation.  In training, the granulation time series can either contain the modal oscillation component or not, 
but, as discussed in the Appendix~\ref{sec:appendixB}, we construct the training set from the three components separately.  

We prepare a $N_t\times Nx\times Ny=40 \times 80 \times 80$ local source response template using the Green's function solution of the propagating wave in two dimensions,
\begin{equation}
\begin{split}
G(x,y,t;x^\prime,&y^\prime,t^\prime) = \\
&{{c_s}\over{2 \pi \sqrt{t^2 - {{x^2 +y^2}\over{c_s^2}}}}}\ H\!\left(t - {{\sqrt{x^2 + y^2}}\over{c_s}} \right)\ ,
\end{split}
\label{eqn1}
\end{equation}
where $H$ is the Heaviside step-function, $c_s$ is the speed of sound, and $x$, $y$, and $t$ on the right hand side are measured relative to the impulse location and time (equal to $x-x^\prime$,  $y-y^\prime$, and  $t-t^\prime$ respectively).  We note that this is not the true Green's function of the three dimensional stratified atmosphere~\citep[e.g.,][]{1998ApJ...496..527R}, but approximates it in the plane of the source height.  We anticipate employing the true Green's function when identifying sources in real data, as that will allow simultaneous extraction of the source height and position, but we use the simplified Green's function here to illustrate the analysis techniques we have developed.  
With Equation~\ref{eqn1} the source response template can be readily constructed as 
\begin{equation}
\phi(x,y,t) =  \int_V G(x, y, t; x^\prime, y^\prime, t^\prime) S(x^\prime, y^\prime, t^\prime)\ dx^\prime\,dy^\prime\,dt^\prime \ .
\label{eqn2}
\end{equation}
%which is a simple three-dimensional (spatio-temporal) convolution.  
Taking $S(x^\prime, y^\prime, t^\prime)$ to be a narrow Gaussian in space and time 
(we take $\sigma_x = 16$ km and $\sigma_t = 2$ so that it corresponds to an unresolved $\delta$-function at the Nyquist frequency of the spatio-temporal grid), 
$\phi(x,y,t)$ serves as the acoustic source response template.

Since the simulated photosphere itself likely has sites of acoustic emission (not a priori identifiable), the MURaM time-series itself cannot be used directly in training the network, as the goal of training is to separate the sources from the other flow components. Instead, using the MURaM photospheric slices, we construct an artificial dataset that captures a source free version of the granulation and its evolution (detailed in Appendix~\ref{sec:appendixB}).  Half of these artificial granulation time-series are used as source free examples for the convolutional neural network, while the other half additionally contain acoustic pulses following the Green's function template above. The acoustic pulses are added to the source free time-series at random positions in space and time, have an amplitude specified by a signal-to-noise ratio (SNR, the ratio between the peak velocity of the acoustic response and the granular flow field at the local site of interest), and propagate at 8 km/s, approximately the mean soundspeed in the simulation photosphere ($\tau = 1$).  

Using these samples, the convolutional neural network is trained to classify a given sample as containing acoustic emission or not; it is trained to determine whether a source is found at a given place and time or not.  
Effectively, the training determines both the connectivity between the network layers and the properties of the series of large and small convolutional kernels applied at each layer so that the loss function is minimized for a given source SNR.

\begin{figure}[t!]
\vskip 0.05in   
\centerline{\includegraphics[width=8cm,trim=0.0cm 0.5cm 0.0cm 0.0cm]{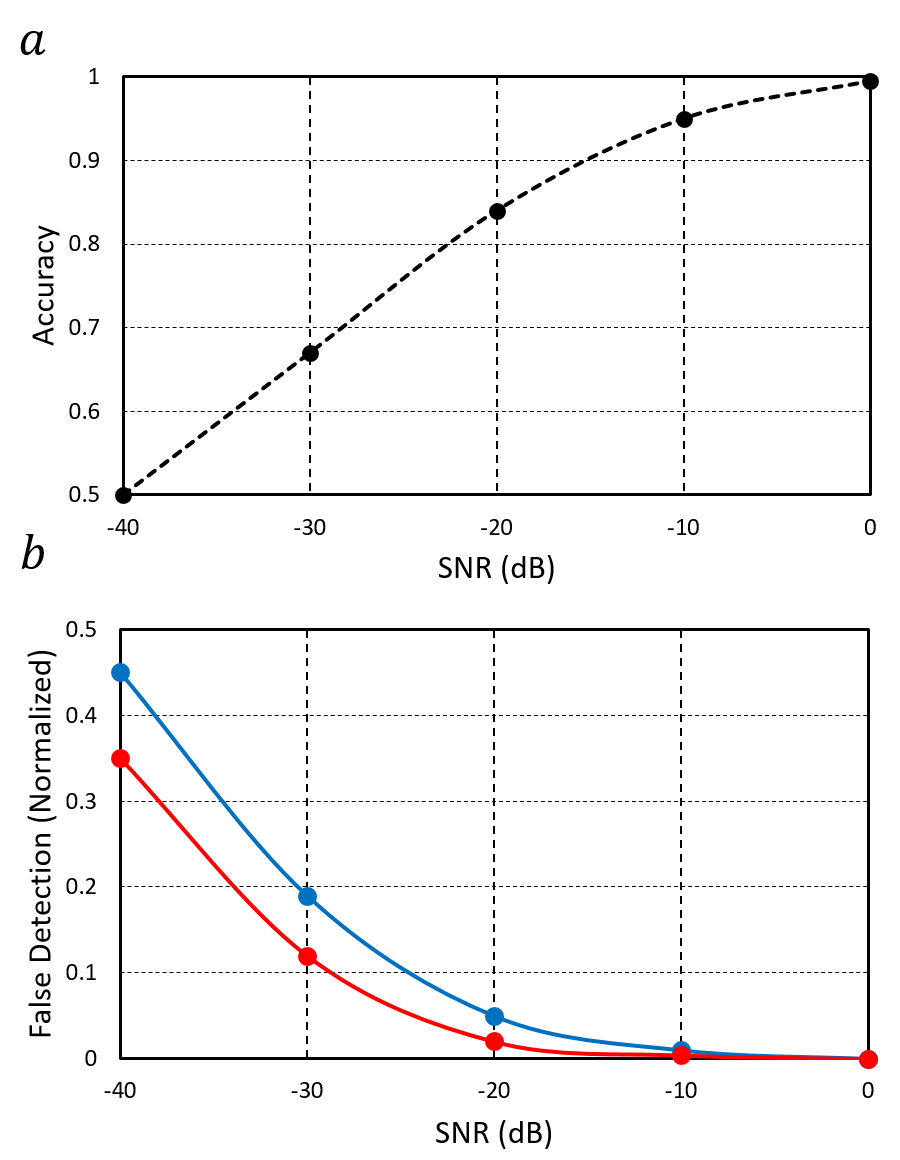}}
\caption{In ($a$), source identification accuracy during network training, one minus the mean-absolute-error over one-thousand test samples (not used in training, but constructed with the same source strength as the training set) as a function of the source strength (acoustic pulse signal-to-noise ratio (SNR) in decibels (dB)). In ($b$), measured rates of false detection after training with -20 dB sources.  Plots separately for false positives ({\it blue/top}) and false negatives ({\it red/bottom}).}
\label{fig:NN-accuracy}
\end{figure}

\begin{figure*}[t!]
\centering
\includegraphics[width=0.75\textwidth]{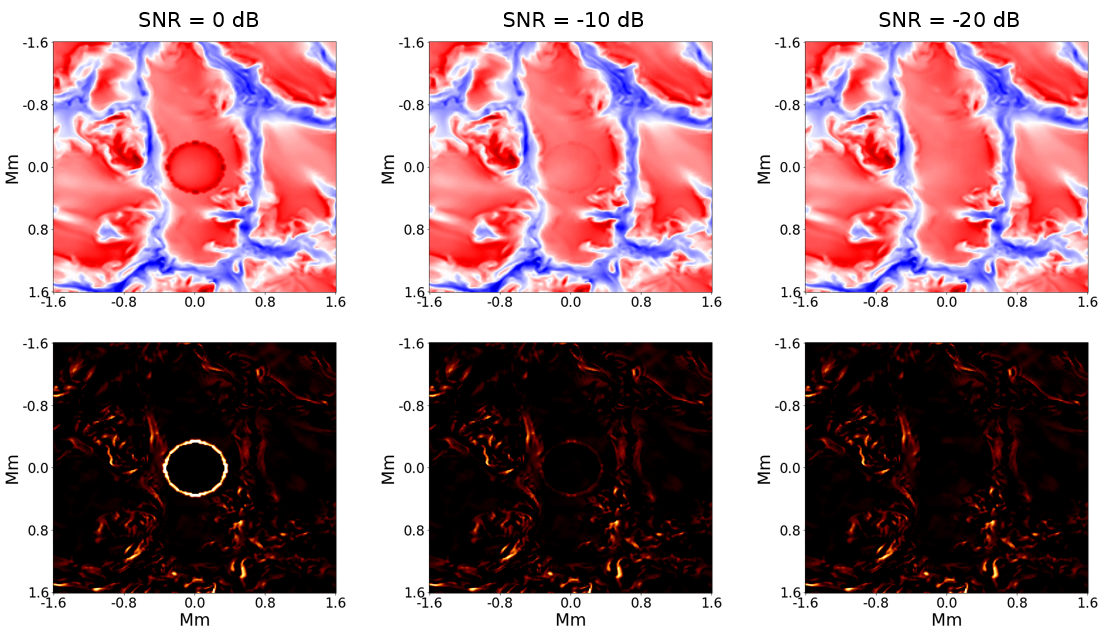}
\caption{Instantaneous visibility of an artificial acoustic pulse as a function of signal-to-noise ratio (SNR in dB) in Doppler map ({\it top}) and in a first difference image ({\it bottom}).  Even though it is located in the a region of relatively uniform granular flow and is centered on the numerical grid, the -10 dB source ({\it center}) is nearly invisible after propagating less than the width of one granule (20 time steps, $\sim41$ s).}
\label{fig:SNR}
\end{figure*}

To assess the best training strategy, we trained the convolutional neural network multiple times, each time with a fixed acoustic source amplitude.  
Figure~\ref{fig:NN-accuracy} ({\it top}) plots the training accuracy for acoustic pulse amplitudes ${\rm SNR}\in\{-10, -20, -30, -40\}$~dB, when the length of the convolutional kernels in time ($n_t$, as detailed in Appendix~\ref{sec:appendixA}) are varied until maximum accuracy is obtained for each source SNR.
Accuracy is defined as one minus the mean-absolute-error, the mean value over a thousand test samples (not used during training) of the source probability as returned by the network after training minus the ground truth.  It is computed  over both source-free time-series and time-series with sources, so that both false positive and false negative detections are accounted for. 
When source SNR equals 0 dB (i.e., the ratio of the maximum amplitude of the acoustic response to the local granular flow is 1 in the training set),
the local wave signal is clearly apparent in individual images (Figure~\ref{fig:SNR}), the loss function of the network converges to a minimum even for $n_t = 3$, and the network can reliably classify the existence of the acoustic emission with an accuracy of $99.5\%$. 
For SNR of -10 dB, the loss function of the network still converges with $n_t = 3$, but the network exhibits a reduced accuracy of $95\%$. 
The accuracy further drops as the SNR of the training source is decreased to -20 dB.  For these weak sources, most of the prominent signatures of the acoustic emission in individual image frames is lost in the granular flow (Figure~\ref{fig:SNR}) and learning convergence is difficult to achieve, requiring longer duration convolutional kernels ($n_t>4$) and multiple training initiations for successful minimization of the loss function. 

The increase in $n_t$ required for learning convergence, which accompanies the decrease in source strength, suggests that at low SNR, convergence requires a noise-specific de-noising filtering that is not accessible in shorter time series but that can be reliably leveraged to allow source identification with longer convolutional kernels.  In other words, as the source strength decreases, the network leverages, via longer duration convolutional kernels, the difference between the temporal evolution of the granulation (the noise) and that of the local wave (its propagation at the sound speed) to identify the source site.  But there is a limit to the source amplitude below which this strategy no longer works:  training accuracy drops to $50\%$ (no better than random chance) for source SNR lower than about -40dB, even when the convolutional kernel includes many time steps ($n_t \gtrsim 10$).  Since amplitude of the wave pulse drops with time as it expands, it longer contributes significantly at these longer times to the characteristic spatiotemporal signature underlying the neural network identification strategy.   

\begin{figure}[t!]
\vskip 0.05in 
%\centering
\includegraphics[width=\columnwidth]{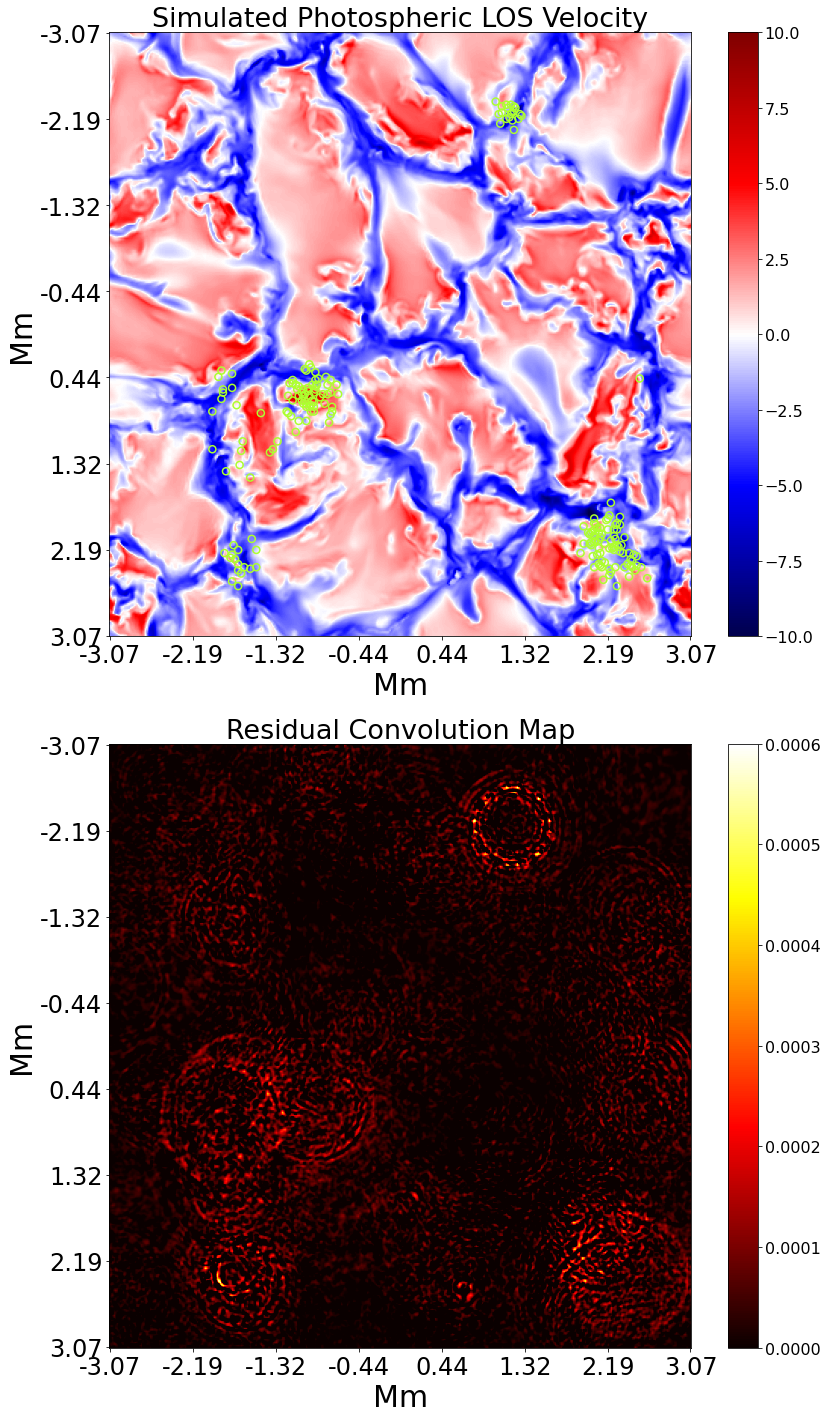}
\caption{({\it Top}) Acoustic source locations found by the convolutional neural network in a single snapshot of the MURaM simulation. Green circles indicate that  the confidence of the algorithm that these are source sites lies above 84\%. ({\it Bottom}) Residual convolutional map resulting from direct application of the filter derived from the temporal kernel used by deep learning algorithm, when applied to the same MURaM photospheric snapshot.  Velocities in the both maps are given in km/s.}
\label{fig:NN-filter}
\end{figure}

Based on this analysis, we trained the final neural network with artificial time series in which granulation and sources have a fixed SNR of -20 dB.  We prepared 5,000 training time series, half with the acoustic pulses randomly located somewhere in space and time, and the other half with no pulse. With these, an ensemble of neural networks, each with different initiation parameters, was trained. 
The network with the highest accuracy on test samples was used in the analysis of the original MURaM data (detailed results in \S\ref{sec:result}). Application of the network to non-training data returns a confidence value at every location in the each image at each time in the time series.  This is effectively a measure of the cross-correlation between the image time series and the Green's function response kernel, but is not a direct measure of that correlation.  As discussed above, the neural network simultaneously applies a de-noising scheme that allows it to recover the spatiotemporal structure of sources underlying the granulation that cannot be easily uncovered otherwise. Figure~\ref{fig:NN-filter} ({\it top}) indicates all the locations in a single time-step at which the neural network returned a confidence value greater than 84\% (indicating 84\% or better confidence that an acoustic source occurred at that location at that time). We note that the network might register multiple source detection as it scans through a single source in space and time. This is because the network can return a confidence value above 84\% even if the source is not precisely at the center of its field-of-view.  Such assignments can lead to multiple false detections, as there can be cases in which confidence of the network at the site of acoustic emission is nearly unity while the confidence of the nearby pixels (in both space and time) remains above 84\%.  The network has finite resolution. From our analysis, we find that a site of strong acoustic emission can cause expanded detection with a spread of about $\pm 6$ pixels in space and $\pm 4$ pixels in time, away from the center of its actual spatiotemporal location (for reference the green symbol in  Figure~\ref{fig:NN-filter} has a diameter of 8 pixels). The clusters in the  field-of-view of Figure~\ref{fig:NN-filter} ({\it top}) indeed contain several sources, densely packed together, each contributing multiple detections.  This is evident in  Figure~\ref{fig:NN-filter} ({\it bottom}), as discussed in the next section.

\begin{figure}[t!]
\vskip 0.05in 
%\centering
%\includegraphics[width=0.7\textwidth]{Figure_NNarch.png}
\centerline{\includegraphics[width=8cm,trim=0.0cm 0.5cm 0.0cm 0.0cm]{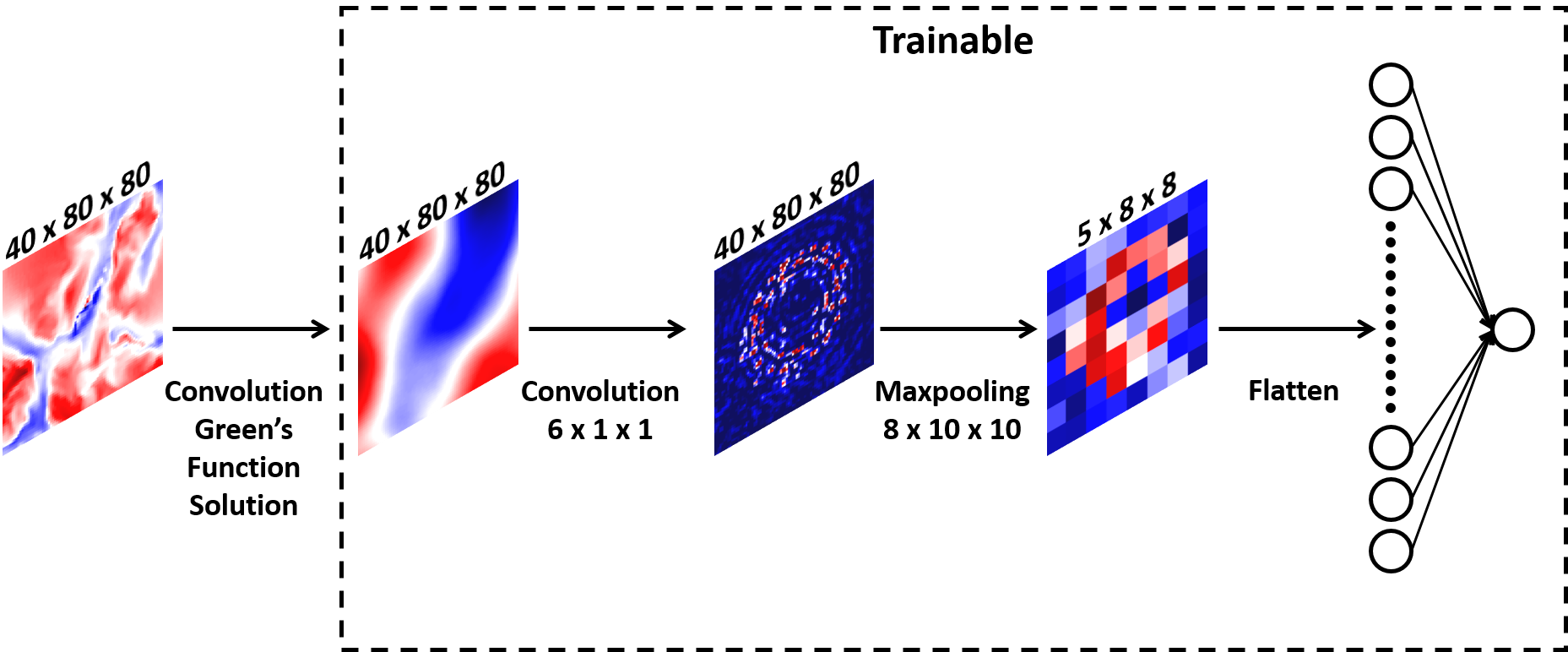}}
\caption{Simplified convolutional neural network designed to extract the de-noising temporal filter that the full network utilized to efficiently identify acoustic emission.}
\label{fig:NN-arch}
\end{figure}

We note that even though the network was trained for sources with fixed SNR of -20 dB, it is able to uncover sources with a range of strengths, depending on their location and phasing with respect to the background granular flow.  This is illustrated by Figure~\ref{fig:NN-accuracy} ({\it bottom}), which plots the fraction of false positives and false negatives (compared to total number samples tested) returned as a function of source strength, when the -20 dB SNR trained neural network is applied to a set of time-series created by embedding one-thousand artificial sources of a given strength at random positions and times in artificial source free granulation time-series.  The network is able to identify $69\%$ of sources of strength -30dB with a false positive rate of $19\%$ and false negative rate of $12\%$.

\begin{figure*}
\centering
\includegraphics[width=\textwidth]{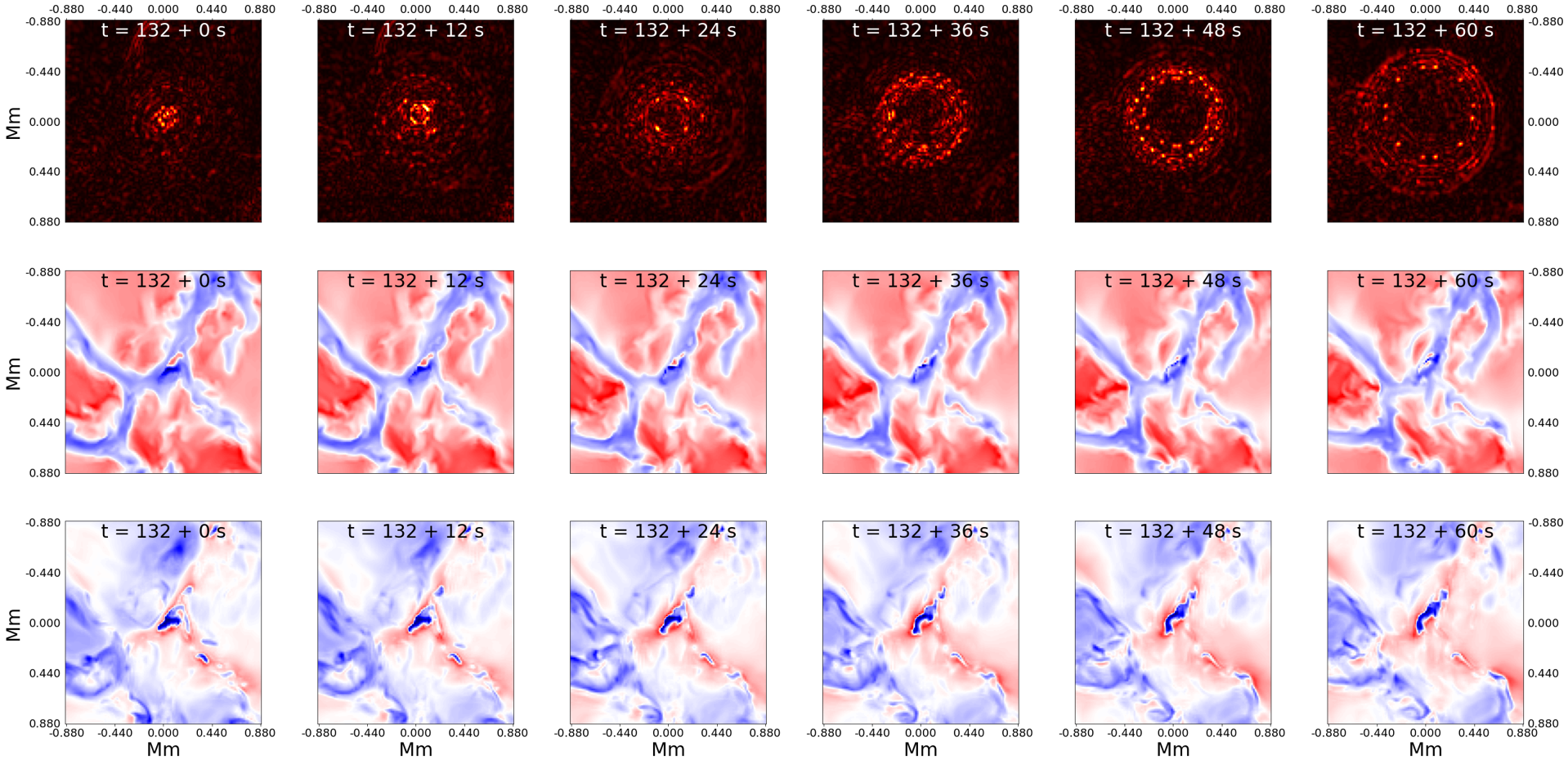}
\caption{Snapshots of residual convolution maps $({\it top})$ illustrating one main impulsive event along with multiple neighboring acoustic wave fronts  propagating at the speed of sound in the MURaM photosphere at $\tau = 1$ plane. The corresponding Doppler velocity field $(center)$ and pressure field $(bottom)$ are shown as well, illustrating the complex dynamics at the source site.}
\label{fig:source}
\end{figure*}

\subsection{Learning-algorithm-derived acoustic-source filter}\label{sec:filter2}

Despite the success of the deep learning algorithm that we have developed, the algorithm itself remains characteristically opaque.  It
is difficult to determine why the network is performing an operation or how it relates to the optimal solution for the problem.  This characteristic opacity is the heart of the ``black box'' problem, a problem with significant practical and theoretical consequences. 
Practically, it is difficult to trust, optimize, and systematically improve an algorithm whose workings are not transparent.  
Theoretically, the black box problem makes it difficult to evaluate the mathematical rigor of the solution and its domain of reliability. 
Additionally, the algorithm only returns the probability that a particular site location is a source.  Alone, this offers limited physical insight. 
To overcome these difficulties, we have unwound the complicated, interlaced convolutional kernels our deep learning algorithm defined, and have deconstructed them into a set of linearly summed traditional operators, converting the ``black box" to a ``glass box."  
Details are provided in the Appendix.  Here we summarize the most salient results. 

\begin{figure*}
\centering
\includegraphics[width=\textwidth]{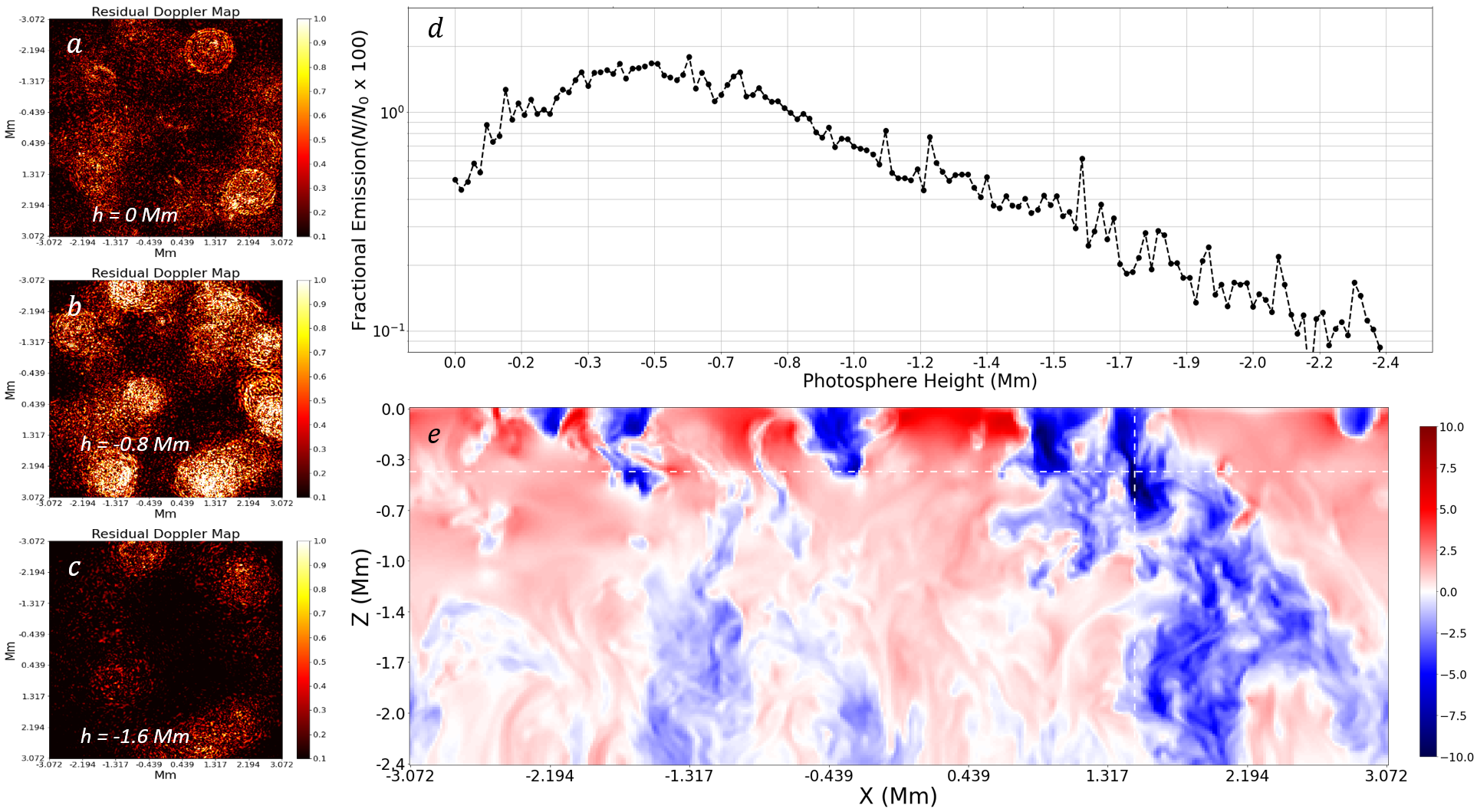}
\caption{$a,b,c$: Variation of acoustic power at different depths in the MURaM simulation. $d$: Fractional emission (in per cent $N/N_0 \times 100 \%$, where $N$ = emission (residual convolution signal) at height $h$ integrated over horizontal plane and time and $N_0$ = emission integrated over in the entire volume) with respect height. The maximum fractional emission occurs at a depth of $400 - 600$ km. $e$: Vertical slice of the  Doppler velocity in the MURaM simulation at the location of the strong source whose evolution is plotted in Figure~\ref{fig:source} (y = -2.194 Mm in the full domain, Figure~\ref{fig:NN-filter}) domain. The {\it white dashed} fiducial lines indicate the $x$ and $z$ event location, the site with the maximum residual convolution signal.}
\label{fig:emission}
\end{figure*}

As described in the previous section, when the source SNR is 0 dB, the spatiotemporal structure of the acoustic emission is prominent in image time-series and the convolutional filter by default concentrates on capturing those geometric patterns in order to identify a source occurrence. However, as the SNR drops below -10 dB, these features are lost in the background (the granular flow field) and the network requires application of a denoising filter to discern the source. From our examination of the neural network behavior in the previous section, it is evident that this de-noising is accomplished by increasing the temporal length of the convolutional kernel. This suggests that the de-noising is taking place along the temporal axis of the data as the signal gets weaker, and based on this understanding, we designed a reduced network aimed at separating the de-noising kernel from the spatiotemporal source kernel.  
Its architecture is sketched in 
Figure~\ref{fig:NN-arch}.  The network begins with a large non-learnable kernel which convolves the input Doppler map with the Green's function solution for the propagating wave (Equation~\ref{eqn1}). This layer serves to capture the spatiotemporal features of the source for the network. The next layer consists of a $6 \times 1 \times 1$  trainable convolutional kernel whose sole purpose is to capture the de-noising scheme essential to the network's success. These convolutional layers are followed by a max-pooling layer which encodes all the information produced by the convolutional filters in a lower-dimensional feature space. This encoded cube is flattened and used by the network to make the decision for identification.

An ensemble of networks with this architecture were trained, using sources having -20 dB SNR as before.  
We examine the trained temporal de-noising kernels achieved and found that they converge to a simple form:  $[0.1 \hspace{2 mm} -0.5 \hspace{2 mm} 1.0 \hspace{2 mm} -1.0 \hspace{2 mm} 0.5 \hspace{2 mm} -0.1]$ (normalized).
This is an oscillatory function, a custom high frequency filter in time, somewhat resembling a Morlet wavelet or a sinc function, but performing better in tests than either of those.  More explicitly, the kernel is the weighted difference of 6 successive planes along the temporal axis of the convolved data cube.  It serves to cancel the background flow, leaving only tiny fluctuations which preserve the residual convolved source response riding on a nearly constant background.  In addition, isolated pixels have very large values, likely caused by granular edges, which can dominate the color table when displaying the images.  To remove these we clip all the large values from the filtered residual timeseries, restricting the residual map to values between $[-0.001 \hspace{2 mm} 0.001]$.  Examples of the results are displayed in Figures~\ref{fig:NN-filter},~\ref{fig:source}, and~\ref{fig:emission}.  Since variations in the residual convolution map this produced do not indicate true upflow and downflows, we plot only its magnitude. 

What is notable is that acoustic sources, and the resulting local propagating wave field they induce, can be directly visualized 
in Doppler map time-series by applying the following neural network motivated operations, in order: 

\begin{itemize}
  \item Convolve the Doppler map timeseries with a template of the Green's function solution for the acoustic response.
  \item Convolve the resulting data cube with the temporal kernel $[0.1 \hspace{2 mm} -0.5 \hspace{2 mm} 1.0 \hspace{2 mm} -1.0 \hspace{2 mm} 0.5 \hspace{2 mm} -0.1]$, or equivalently, apply a weighted difference filter over 6 successive frames along the temporal axis of the convolved data cube.
  \item Clip large values from the filtered timeseries by restricting the residual map to values between $[-0.001 \hspace{2 mm} 0.001]$.
  \item Take the absolute value of the residual map.
\end{itemize}

This procedure can be compared with a carefully defined  Fourier filter in $k-\omega$ space (see Section~\ref{sec:reliability} below). Since it does not depend on the vast set of parameters of the deep learning solution, parameters that are rooted in training constraints, it can be applied as a robust compact mathematical operator directly to observational data.  We are planning to work with early DKIST data to explore that possibility in detail.

\section{Results} \label{sec:result}

Extracting acoustic emission signatures by direct application of the image filter described above (\S\ref{sec:filter2}) to an image time-series has an additional advantage over the neural network.  It allows one to trace the outward propagating wave front, potentially providing more information than strictly the source location, the probability of which is alone provided by the neural network.  This is immediately valuable in distinguishing sources in close proximity.  

Application of the image filter to the photospheric Doppler image time-series of the MURaM simulation reveals that the acoustic sources are frequently found in and near intergranular lanes, particularly at those sites which contain complicated mixed flow structure or sudden local downflow enhancement.  Multiple sources often occur in close proximity.  Figure \ref{fig:source} displays the temporal evolution of the residual correlation map ({\it top} row), Doppler velocity ({\it middle} row), and pressure fluctuations (about the horizontal mean, {\it bottom} row)  in a region with a comparatively isolated strong source.  Even in this case in which one acoustic source is particularly strong, overlapping wave fronts from multiple close-by sources and from somewhat more distant sources can be seen.  These form interference patterns in the residual correlation images. While the wave amplitudes are very low and noise plays some role in the images displayed, we have determined, using artificial and real simulation data and by adjusting the filter applied, that these patterns are not an artifact of the filtering method but instead very likely result from real wave interference. The filtering technique appears to provide a robust method for the identification of acoustic wave fronts emanating from sources that self-consistently arise in the convection simulation.

\begin{figure}[t!]
\vskip 0.05in
%\centering
\includegraphics[width=\columnwidth]{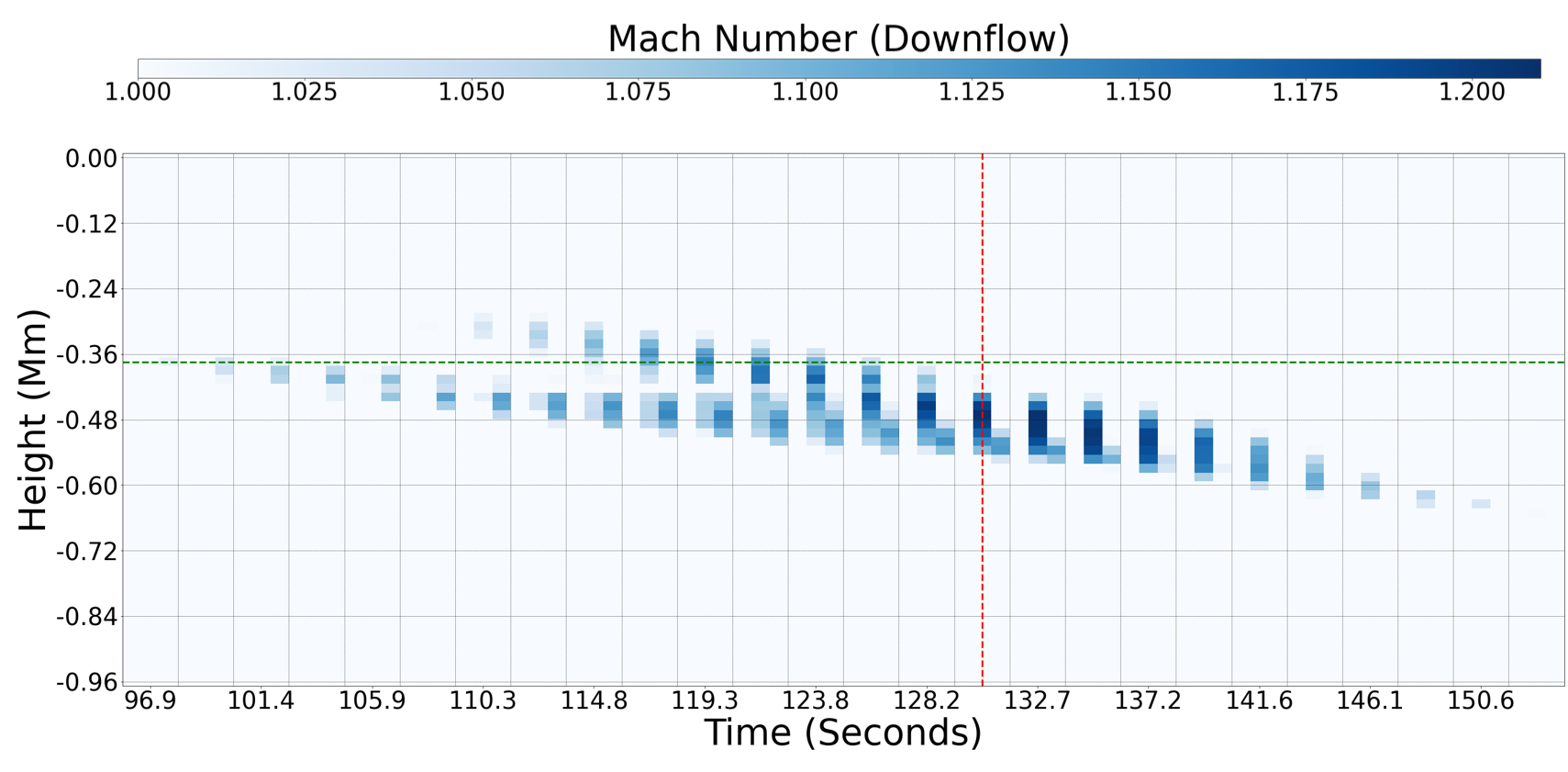}
%\centerline{\includegraphics[width=8cm,trim=0.0cm 0.5cm 0.0cm 0.0cm]{Figure_Mach.png}}
\caption{Descending supersonic downflows plume of supersonic speed as a function of height and time. Each grid box on the horizontal axis indicates a single timestep $\Delta T = 2.0625 s$, with each containing 3-pixels (vertical gridlines demarking three pixels) in $x$ at y = -2.194 Mm. Two flows accelerate and  merge at the spatiotemporal location of the onset of the peak acoustic emission (marked with the {\it red} vertical line).  The {\it green} horizontal line indicates the minimum of the adiabatic exponent, $\Gamma_1$, averaged over the full horizontal extent of the domain.}
\label{fig:supersonic}
\end{figure}

The sources are highly clustered on larger scales as well (see Figures~\ref{fig:NN-filter}, and~\ref{fig:emission}).  There are distinct regions where the acoustic emission is particularly ``loud'' (many sources are found in close spatial and temporal proximity) and others where it is ``quiet'' (few sources).   In the simulation, this structuring appears to occur on mesogranular scales.  To investigate this, we have constructed the residual convolutional map at several heights in the simulation, and have found that acoustic excitation events are clustered around the locations of strong downflows at depth.  The sources appear to be associated with the reconfiguration of the granular flows by deeper convergence of the intergranular plumes into large downflow structures.  The amplitude of the residual convolutional signal is maximum quite deep in the photosphere, with weaker signal both above and below.  
This is visually apparent in Figure~\ref{fig:emission}$a$-$c$, and in Figure~\ref{fig:emission}$d$ we plot the fraction of total signal coming from each height (employing a depth dependent sound-speed in the Green's kernel convolution) integrated over the time series.  The fractional emission ($N/N_0 \times 100$, with $N$ = total residual convolutional signal at given height and $N_0$ = total residual convolutional signal in the volume) peaks at  depth of about $400 - 600$ km below the photosphere, quite deep compared to estimates arising from the study of $p$-mode line asymmetries.

Emission from single very strong events also peaks at these depths indicating that the fractional emission peaks is not just a function of event occurrence rate.  Peak emission (maximum amplitude of the acoustic emission as measured by the residual convolutional signal) from the very strong acoustic event shown in Figure~\ref{fig:source} occurs at a depth of about $480$ km (indicated by the horizontal {\it dashed} fiducial line in Figure~\ref{fig:emission}$e$).  In this case the peak emission occurs as a result of the convergence, within the vigorous mesogranular downflow, of two supersonic granular plumes.  Figure~\ref{fig:supersonic} plots the local Mach number of the flow in a small horizontal slice (3-pixel wide in $x$ and 1-pixel thick in $y$) centered on the vertical dashed line in Figure~\ref{fig:emission}$e$, as a function of time.   Two trans-sonic downflows merge at the position and time of the acoustic event.  
Some previous studies have implicated hydrogen ionization as key to the formation of supersonic downflows and suggested that such downflows play an important role in acoustic excitation~\citep{1993ApJ...408L..53R, 1993ApJ...419..240R, 2001ApJ...561L.191R}.  That seems to be born-out here, with the depth of the minimum of the adiabatic exponent $\Gamma_1=(d{\rm ln}\,P/d{\rm ln}\,\rho)_{\rm ad}$, horizontally-averaged over each depth plane in the simulation, very close to that of maximum acoustic emission ({\it green} horizontal {\it dotted} line in Figure~\ref{fig:supersonic}).      

\section{Reliability Tests} \label{sec:reliability}

Convolutional filters carry some risk that the result one achieves is biased by the convolution one applies, that the pattern one is looking for is accidentally imprinted on the data.  We performed a number of test to help determine if this is the case in our analysis.   In the simplest test, we applied both the neural network and the  convolutional filter to a time series of MURaM photospheric Doppler images after scrambling the phases in time and space (phases randomized over a uniform distribution between zero and two-pi) while preserving the power at each spatial and temporal frequency. The neural network consistently returned a null detection of acoustic emission (confidence values less than 10\%) when applied to this time series, and direct application of the convolutional filter produced some random circular patterns but none that propagated away from a compact site, as does the signal when it is the result of a local source.  This suggest that the convolution is not imposing a defined pattern onto the solution, at least not when the modes are delta-correlated in time and space. 

In another test, we trained the neural network using a particular sound speed in the Green's function source kernel, and then applied it to data samples containing acoustic responses constructed using a range of propagation speeds.  We did this both with and without the granulation noise. The network identified acoustic events with higher confidence when the sample sound speed was similar to that it was previously trained on.  When the sound speed of the test samples deviated significantly from that of the training set, the network returned null detections.  Moreover, when applied to the MURaM simulation data, the networks trained using kernels constructed with a sound speed close to that of the depth being analyzed produced higher confidences (for neural network) or stronger amplitudes (for convolutional filter) than those trained using a significantly different sound speed ($c_s \pm 3$ km/s). Again this suggests that the signal being extracted is in the data, not imposed on it, that the network and the filter are identifying the physical wave response of the medium at the correct sound speed. 

\begin{figure}[t!]
\centering

\includegraphics[width=\columnwidth]{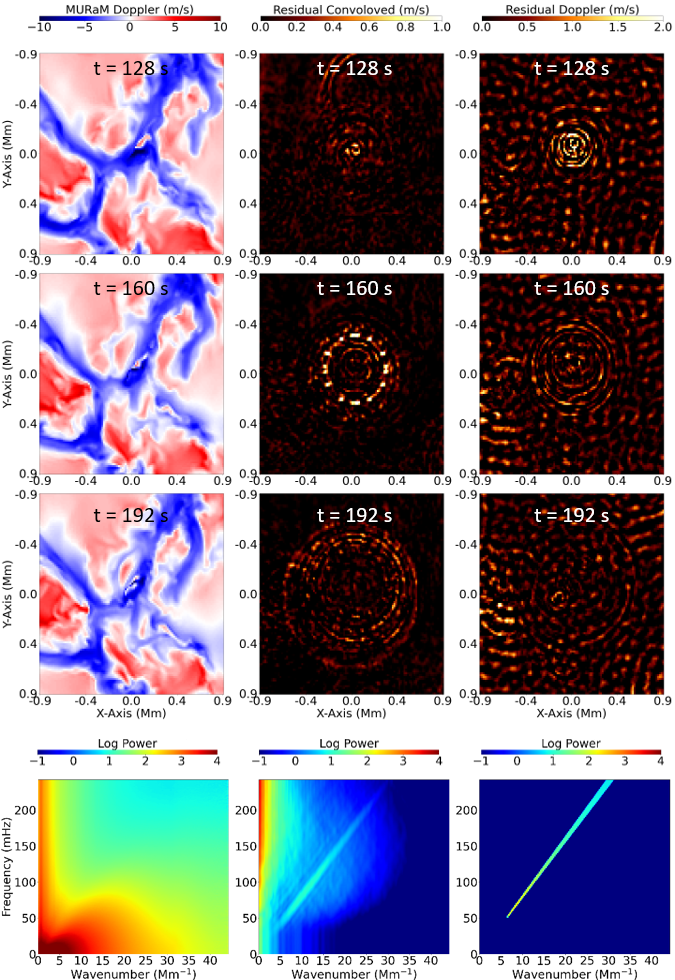}
\caption{Snapshot of the acoustic source in Figure~\ref{fig:source} at $t=128$ s, $t=160$ s, and $t=192$ s. {\it Left} column: Snapshots of the $\tau = 1$ Doppler map from  MURaM.  {\it Center} column:  Residual convolution map.  {\it Right} column: Map of same region after applying simple $k - \omega$ filter that mimicks the convolutional filter to the Doppler image. Botton row shows the corresponding power spectra for image time series from which the column were drawn, full MURaM Doppler map ({\it left}), convolutional filter ({\it middle}) and $k - \omega$ filter ({\it right})}.
\label{fig:k-omega}
\end{figure}

Finally, the neural network based convolutional filter we constructed is effectively a high pass filter, and we explored its characteristics in Fourier space. 
Convolution with the acoustic Green's function followed by application of the high frequency temporal differencing kernel reduces the low-frequency contributions of the granular flow while maintaining Fourier components with phase speeds that lie in the vicinity of the sound-speed (bottom row, middle panel in Figure~\ref{fig:k-omega}).  The filter can be mimicked, to some degree, by constructing a very narrow passband filter on the $k-\omega$ plane, one that filters out everything except high frequency components with phase speeds near the sound speed (bottom row, right hand panel in Figure~\ref{fig:k-omega}).  When this Fourier filter is applied to the data it highlights events very similar to those found using the convolutional filter at the same locations in space and time, albeit with much higher noise levels.  The convolutional filter very effectively extracts from the data those Fourier components with phase speeds near the specified sound-speed.  Those modes have phase relations that corresponded to outward propagating pulses induced by acoustic source events (the Green's function response).

Together these tests provide strong evidence that the convolution operator is not biasing the data to produce local wave-like propagation signals. Neither the neural network nor the direct application of the convolutional filter are prone to finding wave patterns in the granular noise field.   However, one last potential source of contamination can arise.  The intermittent constructive interference of $p$-modes in the data may be identified as a false local source when the  waves lose phase coherence, potentially giving rise to spurious propagating source-like signals as the waves propagate away from the coherence sites.
Fortunately, in these occurrences the modes first come into and then lose coherence.  They can thus be readily identified by their distinct time-reversible signature and eliminated as false sources. We note that the convolutional filter we are applying is dominated by high spatial and temporal frequencies (Figure~\ref{fig:k-omega}), well above those that characterize the solar $p$-modes.  The potential spurious signal we are describing is caused by the interference of the local waves excited by a high density of local sources, not by the global-mode oscillations.   

%Similar results have been achieved using the continuum intensity and pressure fluctuation time series.

%\begin{figure*}
%\centering
%\includegraphics[width=.7\textwidth]{Figure_muram_pmode.png}
%\caption{$k - \omega$ diagrams of a simulated Doppler map of MURaM at $\tau = 1$ (Left). The power of p-mode coherence patches lies at the bottom-left corner of the diagram (showed by white dashed box).}
%\label{fig:pmode}
%\end{figure*}

\section{Conclusion}\label{sec:Conclusion}

We have developed an image time-series filter for the detection of local acoustic perturbations in time-series of photospheric Doppler velocity images.  We have achieved similar results, not described in this paper, using the continuum intensity and pressure fluctuation image time-series. The neural network motivated convolutional filter we have described is quick to apply and can be applied to input images of arbitrary size and time-series of arbitrary duration.  It has no parameters to tune.  Making the interpretive step away from the neural network itself frees us from the need for the very large observational data sets required to train deep learning algorithms. The algorithm we developed relied on an idealized simulated environment.  The machine-learned knowledge was then interpreted in terms of human-understandable operations.  Those the operations can be directly applied to observations without retraining.  Moreover, the interpretability if the filter we have developed allows us to test the range of the filter's applicability and tune it to optimize its sensitivity when applying it to real observations.  It may be possible to improve the current version of the filter via architectural adjustments to learning scheme of the neural network, and the reliability of the filter should be tested over a wider range of simulations, but, given its initial performance on the simulation data as described in this paper, it may likely already be sufficiently robust to make significant contributions upon its initial application to real data. 

Although we focused this work on photospheric source detection, using a two-dimensional approximation to the Green's function, we were able to analyze height-dependent effects in the three-dimensional simulations by adjusting the kernel sound speed to match that of the depth of the layer being analyzed.  Observations are limited in the depth to which they can probe, and so while the  two-dimensional  Green's function may be applied upward in the solar atmosphere, it will not be able to identify the location of deeper sources, those at the depth of peak emission suggested by our work.   
However, photospheric signal of the true three-dimensional Green's function is sensitive to the source depth, and application of similar machine-learning techniques we employ here may allow the determination of the location and depth of source events 
using high-cadence high-resolution Doppler measurements at one or more heights in the observable solar photosphere.     
Fortuitously, the observational capabilities required for these efforts are just now becoming available with the commissioning of the National Science Foundation's Daniel K. Inouye Solar Telescope (DKIST).  

The implications of this work extend beyond identification and characterization of the source of the solar $p$-modes.  As examples, measuring the nonuniform source distribution in the photosphere may lead to an understanding of the spatially inhomogeneous in the propagation of energy and momentum into the chromosphere and consequent observable footprints in the wave flux and power spectra measured there, and the ability to carefully measure the very high spatial and temporal frequency local propagating wave front induced by real sources may open up a new era in high-resolution local helioseismological sounding of small scale structure in the photosphere.  

\section{Acknowledgements}\label{sec:acknowledge}
The authors sincerely thank M. Rempel for providing the MURaM simulation cube and C. Lindsey for noting the time-reversability of the $p$-mode coherence signal. This work was partially supported by National Science Foundation grant number 1616538 and the National Solar Observatory's DKIST Ambassadors program. The National Solar Observatory is a facility of the National Science Foundation operated under Cooperative Support Agreement number AST-1400405.

\bibliography{Manuscript}{}
\bibliographystyle{aasjournal}

\appendix

\section{Neural Network Architecture} \label{sec:appendixA}

The network we employ has a convolutional architecture, which applies a series of convolutions with a combination of small and large kernels (to be inferred during the training) to the input data at each layer.  The neural network was developed using the Keras Python library with the Tensorflow backend, and executed on Dual NVIDIA Quadro P5000 16GB GPUs.  All inputs were normalized to the interval $[-1, 1]$ in the training set. The training was carried out by minimizing the  ``binary crossentropy'' loss function via an Adam stochastic first-order gradient-based optimization algorithm~\citep{2014arXiv1412.6980K} with an adaptive learning rate.  As for any stochastic optimization method, the gradient was estimated from subsets of the input samples, also known as batches. We used batches of 2 samples and trained the network for 50 epochs, thus each training instance runs 12.5 million iterations to convergence.

The network architecture:

\begin{itemize}
      \item Input: this layer represents the input images of size $N_t \times N_x \times N_y$. Consequently, it accepts tensors of $N_t$ image sequences each $N_x \times N_y$ in size.
      
      \item Convolution I, $n_t \times 5 \times 5$: this layer represents four-dimensional convolutions with a set of 64 kernels (channels) $N_{input} \times n_t \times 5 \times 5$ in size. We iteratively determined the number of kernels and their size to provided best inference, with the network still being trained very fast using the GPUs. The output tensors of these layers are $64 \times N_t \times N_x \times N_y$ in size.
      
      \item Maxpool, $1 \times 4 \times 4$: this layer simply down-samples the output from previous layer, reducing its spatial dimensional and allowing for assumptions to be made about features contained in the sub-regions binned.  The output tensors of this layer are $64 \times N_t \times N_x/4 \times N_y$ in size.
      
      \item Convolution II, $n_t \times 3 \times 3$: another layer of four-dimensional convolutions with a set of 32 kernels (channels) of $N_{input} \times n_t \times 3 \times 3$ in size. Again, we iteratively determined the number of kernels and their size to provided best inference within the limits of performance. The output tensors of these layers are $2048 \times Nt \times Nx/4 \times Ny/4$ in size.
      
      \item Maxpool, $1 \times 2 \times 2$: another layer that down-samples the output spatial dimension further, resulting in output tensors $2048 \times Nt \times Nx/8 \times Ny/8$ in size.
      
      \item Flatten: This layer flattens the output from the previous layer to a one dimensional array. Hence, the dimension of the output array of this layer is $(2048 \times 3 \times Nx/8 \times Ny/8, 1)$.
      
      \item Fully Connected, $10 \; neurons$: A fully-connected layer of 10 neurons with tanh activation which implements the operation: $activation(out_{flatten} \cdot W + b)$ where activation is the element-wise activation function passed as the activation argument, $W$ is a weights matrix created by the layer, and $b$ is a bias vector created by the layer.
      
      \item Output, $1 \; neuron$: A single neuron fully connected with the previous layer and activated with soft-max activation to calculate the probability of the target. The range of the output in the neuron is 0 to 1 as this layer returns the confidence of whether an acoustic emission occurs or not.
      
   \end{itemize}
   
\section{Training Set for Neural Network}\label{sec:appendixB}

\begin{figure}[t!]
\centering
\includegraphics[width=\columnwidth]{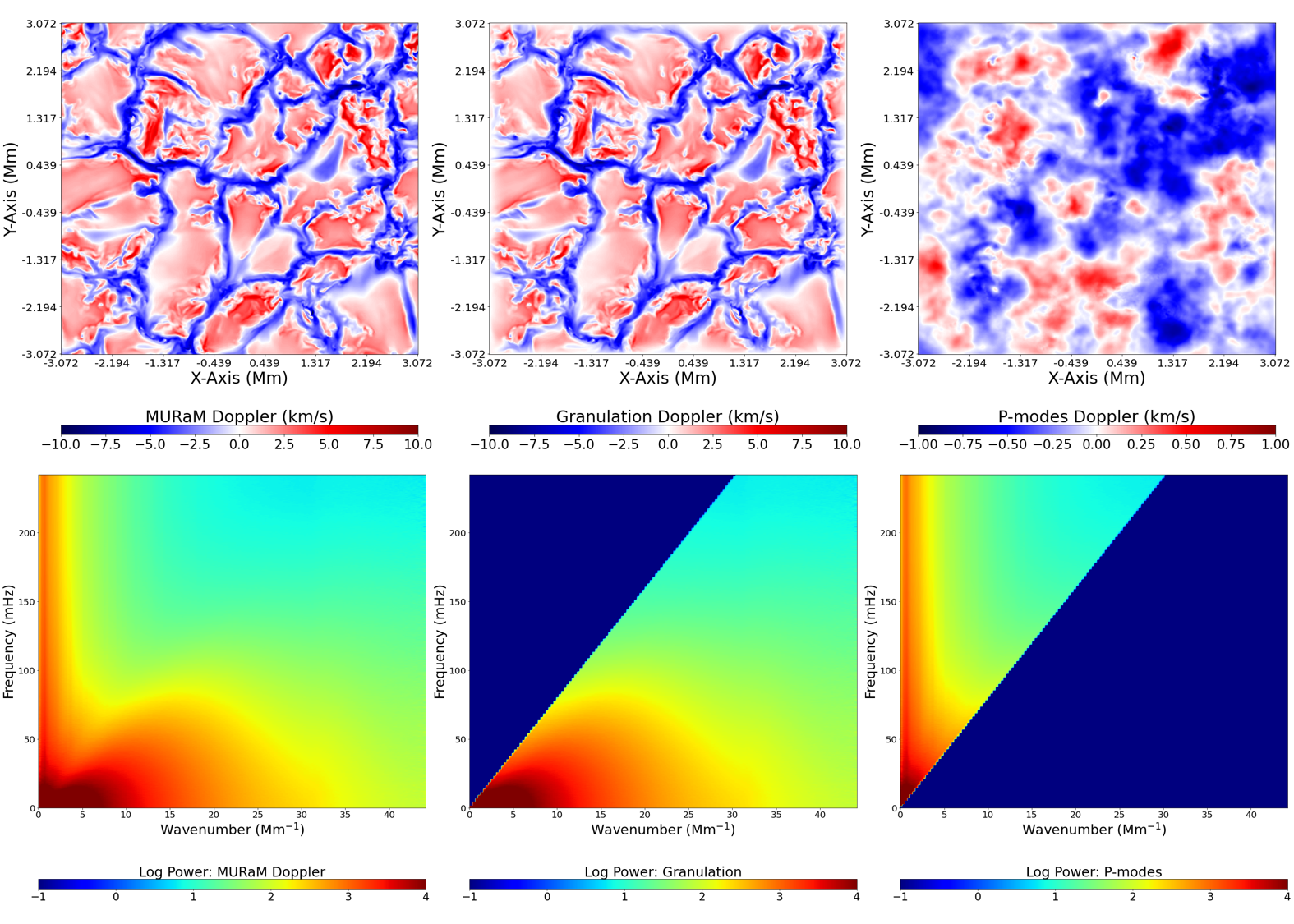}
\caption{Top: MURaM photospheric Doppler map $(right)$, filtered granular motion $(center)$ and filtered modal oscillations $(left)$. Bottom: Corresponding $k-\omega$ diagrams.}
\label{fig:app_kw}
\end{figure}

The MURaM photospheric data can be thought of as being composed of three intrinsic components, where the two of these are dominant, the convective motions and the modal oscillations, and the third is faint, the wave field produced by local sources. The dominant components are shown in Figure~\ref{fig:app_kw}. Since the simulated photosphere itself has sites of acoustic emission, the data  sed for training the neural network needs to be sanitized in such a way to diminish the contribution of the acoustic emission events. 

One way to achieve this is to filter the MURaM photospheric data 2 km/s (as determined by empirical testing on idealized sources) below the sound speed limit of the typical subsonic filter~\citep{1989ApJ...336..475T} leaving only the granular motion, and then adding a random mixture of all the allowed modal oscillations in the simulation box. The resulting composite Doppler map includes only very limited contributions from any source induces acoustic pulse as its Fourier contribution is concentrated along the constant phase speed line in the $k$-$\omega$ diagram. It can be used as source-free template.

Similar result can be achieved using deep learning algorithms as well. We experiment with a convolutional variational auto-encoder, which is essentially a generative model that learns separately the granular motion and the modal oscillations. For such cases, two datasets one filtered 2 km/s below the sound speed limit and one filtered 2 km/s above (such as Figure~\ref{fig:app_kw} {\it right} panel) are prepared and used to train two individual generative autoencoders. The generated granular motion and the modal oscillations are then mixed with appropriate amplitudes and the final training data is produced. This composite Doppler map is found to be predominantly source-free and can be used as source-free template as well. 

We explored the performance of augmented datasets produced by both methods described above, and achieved similar outcomes, concluding that both methods are equally viable.




\end{document}